# Preview, Accept or Discard?
# A predictive low-motion interaction paradigm

Jose Berengueres

Nazarbayev University & KTH Royal Institute of Technology

jose.berengueres@nu.edu.kz

Repetitive strain injury (RSI) affects roughly one in five computer users and remains largely unresolved despite decades of ergonomic mouse redesign. All such devices share a fundamental limitation: they still require fine-motor motion to operate. This work investigates whether predictive, AI-assisted input can reduce that motion by replacing physical pointing with ranked on-screen suggestions. To preserve user agency, we introduce Preview–Accept–Discard (PAD), a zero-click interaction paradigm that lets users preview predicted GUI targets, cycle through a small set of ranked alternatives, and accept or discard them via key-release timing. We evaluate PAD in two settings: a browser-based email client and a ISO 9241-9 keyboard-prediction task under varying top-3 accuracies. Across both studies, PAD substantially reduces hand motion relative to trackpad use while maintaining comparable task times with the trackpad only when accuracies are similar to those of the best spell-checkers.

## 1 INTRODUCTION

### 1.1 Prevalence and Persistence of RSI

Computer users perform millions of clicks each year—up to 3 million for gamers and about 1 million for typical office workers. Around 20% of them report chronic discomfort or repetitive strain injury (RSI) symptoms related to prolonged PC use—an exposure that combines sitting posture, keyboard use, and mouse operation. These numbers were confirmed by our own independent survey, where from a sample of 28 students of age under 30, four reported having a diagnosed RSI related to mouse use [9]. These rates place RSI among the most prevalent occupational-health conditions worldwide—a scale often described by some authors as epidemic in scale [3, 29, 33]. Several efforts have attempted to improve the situation and can be classified in two buckets (hardware and software). One way is to improve the mouse (e.g. slanted mouse). The other route is to rationalize the pointing task—usually via software, (e.g. bubble cursor, pointer acceleration settings).

### 1.2 Limits of Physical Ergonomics

These hardware approaches include numerous ergonomic mouse redesigns proposed over the past decades. Figure 1 illustrates a recent example from our group [7]. However, while partially effective, all ergonomic mice share an intrinsic flaw: they still require physical movement to operate—the root cause of RSIs. Here, we ask whether predictive AI can reduce those movements. In particular, we explore how to replace physical point-and-click motion with AI-generated predictions, thereby trading-off trackpad/mouse fine motor control effort (Fitts's Law) for effort to select suggestions (Hick's Law). However this trade-off is not absolute but becomes hybrid. Between 1980 and 2000 several studies characterized such composite cost of graphical-user-interface selection as the sum of decision and pointing components.

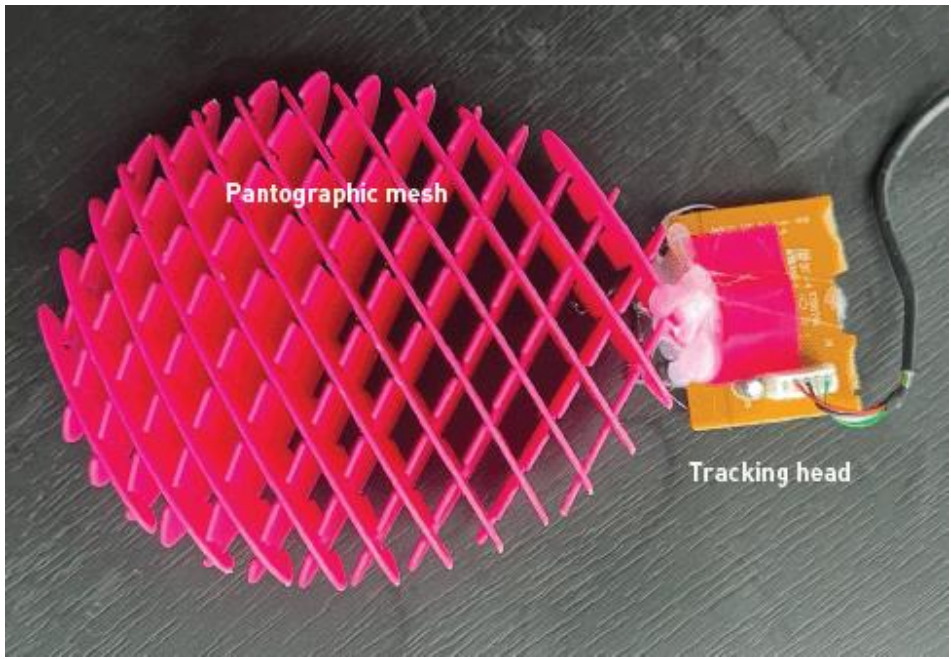

Figure 1: Has progress of physical-based ergonomics plateaued? In the photo a Fleximouse prototype. Source: ACM Interactions.

### 1.1 Hybrid models

Landauer and Nachbar (1985) formalized graphical-menu selection time as the sum of decision and pointing components—time = decision cost + pointing cost tasks [22]. Cockburn et al. (2007) later demonstrated that a hybrid model combining Hick's and Fitts's laws predicts GUI selection performance more accurately than either law alone [11]. Soukoreff and MacKenzie (2004) showed in text-entry experiments that the Fitts-law *slope* remains constant across input methods, with efficiency differences captured mainly by intercept shifts [32].

However, the role of prediction in such hybrid systems remains uncertain. What happens when we integrate AI-assisted prediction into the interaction loop? As shown later in this paper, even when predictive accuracy reaches 100 % within a Top-3 ranking, these hybrid effects persist. This observation aligns with prior findings that automation can reduce—but not eliminate—motor and cognitive control demands. Predictive pointing and touch systems continue to exhibit Fitts-like feedback loops despite anticipatory algorithms [1, 15, 18, 28, 31], and predictive text-entry techniques likewise trade keystroke savings for additional decision effort [20].

### 1.2 Preserving Agency in AI Interaction

Nevertheless, if AI is to assist, agency must be preserved as it is a central concept in human–AI interaction design [2]. An acceptance mechanism is needed, one that is both low in RSI risk and low in cognitive load. At minimum, users should be able to control how to accept or reject an AI-generated click suggestion intended to replace a traditional point-and-click. Ideally, the interface should also provide a preview mechanism to help users verify intent and avoid the all-too-familiar "Reply All" mishap.

Early attempts to operationalize such mechanisms can be traced to the *One-Press* system, which embedded navigation, preview, and commit/abort functions within a single press-and-release cycle on a pressure-sensitive keyboard[12]. *One-Press* exemplified how agency and low cognitive friction could coexist: the user's intent was continuously previewed and confirmed through the press-hold-release gesture. Its design later inspired features such as Spotlight Search and the macOS application switcher, both of which rely on lightweight preview-before-commit action. However, *One-Press* remained hardware-specific and was never adapted to standard web or DOM environments.

### 1.3 Preview, Accept or Discard

To address this gap, we introduce a mechanism that preserves user agency when interacting with AI-predicted point-and-click targets. This mechanism extends the *One-Press* concept by adding the ability to cycle through multiple predicted options and by generalizing the interaction to any standard keyboard. We term this paradigm Preview–Accept–Discard (PAD).

The novelty of PAD lies in unifying two established interaction techniques: (i) cycling through alternatives [25]—such as toggling with the spacebar, as in the macOS application switcher—and (ii) overloading key functions via release-timing control [36]. These techniques are here unified within a single **low-motion interaction grammar**. Figure 2 illustrates an email-client mock-up implementing PAD. Figure 3 presents a ISO 9241-9 test screen with PAD enabled.

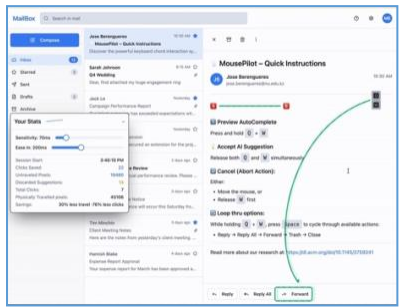
a. Preview of a suggestion (Q+W)

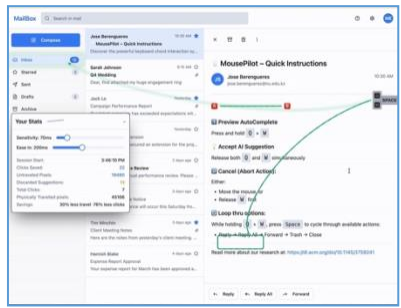
b. Transition to next (Q+W+Space)

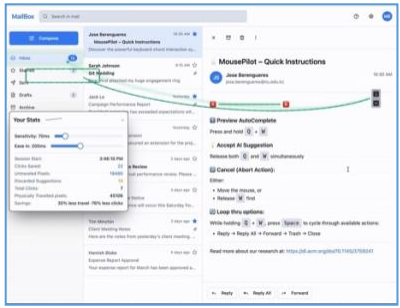
c. Transition underway (Q+W)

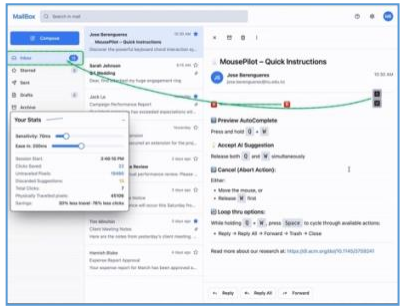
d. Transition completed (Q+W)

Figure 2: An email client mockup that used Q+W as shortcuts with emerald-color chords. The chord points to the most likely ranked clickable target and will execute click on user "accept".

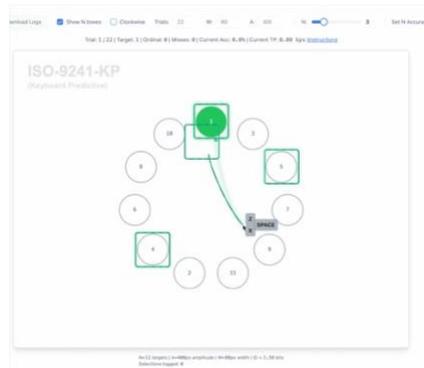

Figure 3: ISO 9241-9 is typically used to compare point-and-click performance not ergonomics. Here a version of same is adapted for Keyboard-Predictive mixed input method. Online at https://ai-mouse.lovable.app/

From an RSI perspective, the PAD can be viewed as an attempt to replace the mouse with the keyboard, under the assumption that the latter entails lower RSI risk. Note that this assumption is currently only partially supported by empirical evidence[19]. In addition, empirical data about carpal tunnel surgery do not show an overwhelming imbalance towards the dominant hand that operates the mouse or trackpad (usually the right hand) but rather just a slight difference. For example, in a local hospital in Astana the CTS surgery intervention split was 45.3% Right hand only, 32.8% Left only, 21.9% both (N=60). This R:L ratio is 1.21 CI 95% [0.77,1.96], meaning we cannot rule out no difference at 95% level.

On the other hand, from a keyboard use perspective, PAD can be seen as a universal shortcut that lessens the cognitive load—reducing the frequency of *homing*. In other words, the shortcuts keys are always the same, what they do is an AI-guess of the user's intent, thereby alleviating the main limitation of conventional shortcuts—users often fail to remember them [24].

Taken together, these ergonomic and cognitive viewpoints—combined with the increasing commoditization of predictive AI—expose a clear gap in the literature: the absence of a unified framework for AI-assisted, low-motion interfaces that preserve user agency while minimizing both cognitive load and RSI risk. In the following sections, we lay out such a framework and use it to evaluate the proposed PAD in two settings.

## 2 BACKGROUND

### 2.1 The Ergonomics of Low-Motion Interaction

Efforts to mitigate repetitive-strain injuries (RSI) in computer input have produced a wide range of ergonomic mice—vertical, slanted, trackball, [26, 30] track bar, biofeedback-based, mouse trap, flexible [7] and even power-assisted [8]. Although such designs are reported to be partially effective, none is 100% effective as they all share a fundamental limitation: the need for continuous fine-motor motion to operate. One work around is to replace the mouse with the keyboard. Keyboard-first paradigms promise reduced homing (as a switching cost between mouse and keyboard) and less hand motion, suggesting a path toward decreased RSI risk. Conversely, predictive interfaces propose actions on behalf of users, introducing a mixed-initiative form of collaboration between human and system.

### 2.2 Predictive and Mixed-Initiative Input

Early research by [10] demonstrated online adaptive **gesture** prediction for command selection—one of the first instances of predictive assistance in input control. However, it lacked inline previews or explicit user acceptance phases necessary to maintain user agency. Later predictive-menu and autocomplete systems incorporated **adaptive ranking** [4, 14], yet none combined real-time preview with lightweight confirmation UI/UX. As we will see later, PAD builds directly on this trajectory by coupling prediction with explicit user control through timing-based acceptance.

## 2.3 Press–Hold–Preview–Release Cycles

The most direct conceptual predecessor is the aforementioned *One-Press* from 2010, which embedded navigation, preview, and commit or abort within a single press-and-release cycle on a pressure-sensitive keyboard. It introduced the WYTIWYG ("What You Touch Is What You Get") principle—inline feedback followed by immediate acceptance on key release. Although innovative, its reliance on pressure sensing limited adoption, and no web/DOM-specific implementation was explored. (The Chrome Web Browser Extension Store only launched later in 2010). PAD generalizes this press–hold–preview–release idea to commodity keyboards via release-timing detection and provides a mechanism for candidate cycling (useful when the AI top-1 accuracy <100%).

## 2.4 Keyboard-First Action Pointing

Beyond *One-Press*, several commercial systems have explored keyboard-centric pointing. Commercial systems such as *ModeKeys* and *AimKeys* [6] reconfigured command activation around keyboard-only triggers, eliminating the need for mouse movement and showing both faster and preferred performance across multiple tasks. They confirmed the ergonomic and cognitive advantages of homing-free operation. Similarly, other systems such as *KeySlide* [27] used finger-slide gestures on physical keyboards for **cursor** control but remained gesture-based rather than predictive or preview-driven. PAD instead performs disambiguation between several candidate targets by cycling via the spacebar. Conversely, smartphone "word gesture" keyboards (now adopted in many smartphones), building on shorthand recognition ideas, brought gestures and chord like path previews to the keyboard, approximating typing by sliding between keys [35]; PAD "inverts" these features by bringing prediction and chord like preview from the keyboard to the screen in a desktop/laptop context.

## 2.5 Guidance and Shortcut Learning

Learning and recall of large numbers of shortcuts have long challenged users. Finger-aware shortcuts and guided variants *FingerArc* and *FingerChord* [36] demonstrated that showing visual guidance during key holds improves speed and reduces errors. *KeyMap* [24] further showed that the timing of on-hold guidance influences expert flow. *KeyMap*, in particular, visualized shortcuts directly on a virtual keyboard, improving recall and discoverability in web contexts. Regardless of their effectiveness, these works collectively show that humans are poor at memorizing large shortcut sets, motivating visual and temporal aids. In the same vein, [4] introduced *Transition*, a model describing how users evolve from novice to expert interaction over time. This emphasized that adoption depends on two key measures: perceived **benefit** and low cognitive **friction**.

## 2.6 Research Gap

In summary, the aforementioned prior works separately contribute essential components: inline preview within a press–release cycle [12]; keyboard-triggered action pointing (*ModeKeys*, *AimKeys* [6]); and guided discovery for shortcut learning (Finger-Aware, *KeyMap* [24, 36]). Yet no existing work combines these capabilities with predictive target **ranking**, bounded candidate **cycling**, and **release-timing** acceptance on commodity keyboards.

Furthermore, previous evaluations of new input methods rarely focus on measuring the amount of motion required from the user. Focusing instead on recall, speed and other more poignant performance metrics. However, given the growing scale of RSI incidence among computer users, here we will focus in measuring the both efficiency and motion reduction.

Beyond ergonomics, this work also raises broader HCI questions. If such "smart" pointers become more popular, who controls the prediction loop? Will AI-enabled input devices become embedded in proprietary agentic browsers (Comet, Pilot, Atlas, Strawberry), or evolve into contextual co-agents competing/collaborating with today's omnipresent chat interfaces (Claude, Copilot, Gemini, ChatGPT)?

## 2.7 Research Questions and Hypotheses.

Inspired on prior pointing-facilitation work in two-dimensional selection—such as the Bubble Cursor [17]—we hypothesize that a predictive selection would significantly reduce total pointer travel without degrading perceived control. Accordingly, here we investigate whether the proposed Preview–Accept–Discard interaction can (i) reduce physical hand motion compared with standard mouse or trackpad use, (ii) maintain comparable task-completion time, and (iii) preserve a sense of agency.

# 3 DESIGN

## 3.1 Principles

PAD is guided by three design goals derived from prior literature:

1. Minimize motion by shifting fine-motor pointing effort from the mouse to the keyboard (shortcuts)
2. Preserve agency through explicit, reversible acceptance low friction actions; and
3. Lower cognitive friction by using timing and feedforward preview rather than new modes or widgets

These goals extend the *One-Press* WYTIWYG concept into commodity hardware and situate it within predictive, mixed-initiative GUI control. The design seeks to balance prediction efficiency with human control—an established tension in mixed-initiative systems. Algorithm 1 shows a high level pseudocode that was used to implement it.

```
Algorithm 1. The Preview–Accept–Discard interaction cycle
─────────────────────────────────────────────────────────
On GUI update:
    T ← Predict(DOM)              ▷ rank top-N targets
    i ← 1
While PAD_mode:                   ▷ true if Z+X held down, false if neither key is held down
    PreviewChord(T[i])
    On keyboard update:
        Next:    i←(i mod N)+1    ▷ Z+X + spacebar
        Accept:  ClickOn(T[i])    ▷ release Z+X simultaneously
        Discard: HideChord()      ▷ release Z+X sequentially
end
```

### 3.2 From One-Press to PAD

As mentioned, *One-Press* integrated navigation, preview, and commit or abort within a single keypress cycle on pressure-sensitive keyboards. PAD re-interprets the same interaction grammar for any standard keyboard by encoding commit or discard through release timing between two modifier keys (e.g., Z + X) while using the spacebar for candidate cycling. This preserves the temporal semantics of *press–hold–preview–release* while remaining deployable on commodity devices.

### 3.3 Predictive Mixed-Initiative Control

Whereas *One-Press* operated deterministically, PAD ranks likely targets based on DOM, button size, accent color, label, history, etc., and context. This introduces a **mixed-initiative loop**: the AI proposes → the user previews → accepts, cycles, or discards. Inline previews (chords inspired by those used in 3D/VR [5]) ensure transparency and help the user maintain situational awareness to elicit trust and perceived control [13].

### 3.4 Release-Timing Acceptance

The design adopts a default 170ms differential user-adjustable release window between modifier keys (Z+X) as the semantic boundary between simultaneous and sequential release. Further debounce and hysteresis filters to avoid accidental commits (clicks) were not implemented. All processing occurs on the client side. See also Algorithm 1.

### 3.5 Chord Preview and Candidate Cycling

Predicted targets are displayed as overlay chords. Candidate cycling is intentionally bounded to 2 options in email compose screen and 6 options in the main screen. These numbers are chosen to mirror the typical range of options used in the most common spell checkers which currently exceed 90% top-1 accuracy. However, no optimization or cost–benefit analysis such as that used to evaluate the *Metropolis* keyboard has been attempted [34]. The spacebar cycles through the ranked list.

### 3.6 Summary

In summary, PAD operationalizes four intertwined ideas—prediction, preview, timing, and cycling—into a single, low-motion interaction grammar. By embedding acceptance semantics in key-release timing and providing real-time visual feedback, PAD aims to maintain user agency while avoiding using more than three fingers. This directly informed the system implementation described next. We avoid using typical modifier keys such as CTRL and SHIFT as those tend to conflict with existing applications and the OS.

## 4 IMPLEMENTATION

### 4.1 Prototype Overview

A prototype was implemented using a low-code web hosting platform as a proof-of-concept. It simulates an email client scenario with a static, browser-based mockup. The design goal was to explore feasibility of Preview–Accept–Discard, not to evaluate algorithmic accuracy. For simplicity, the prototype uses hard-coded predictions that are always correct (100 % Top-N Accuracy) across two screens—the main inbox and a secondary email compose view. This allows controlled testing of the interaction independent of AI model variance. Figure 3 shows a six screen path a user might follow when asked to click reply and then click on send in the mockup.

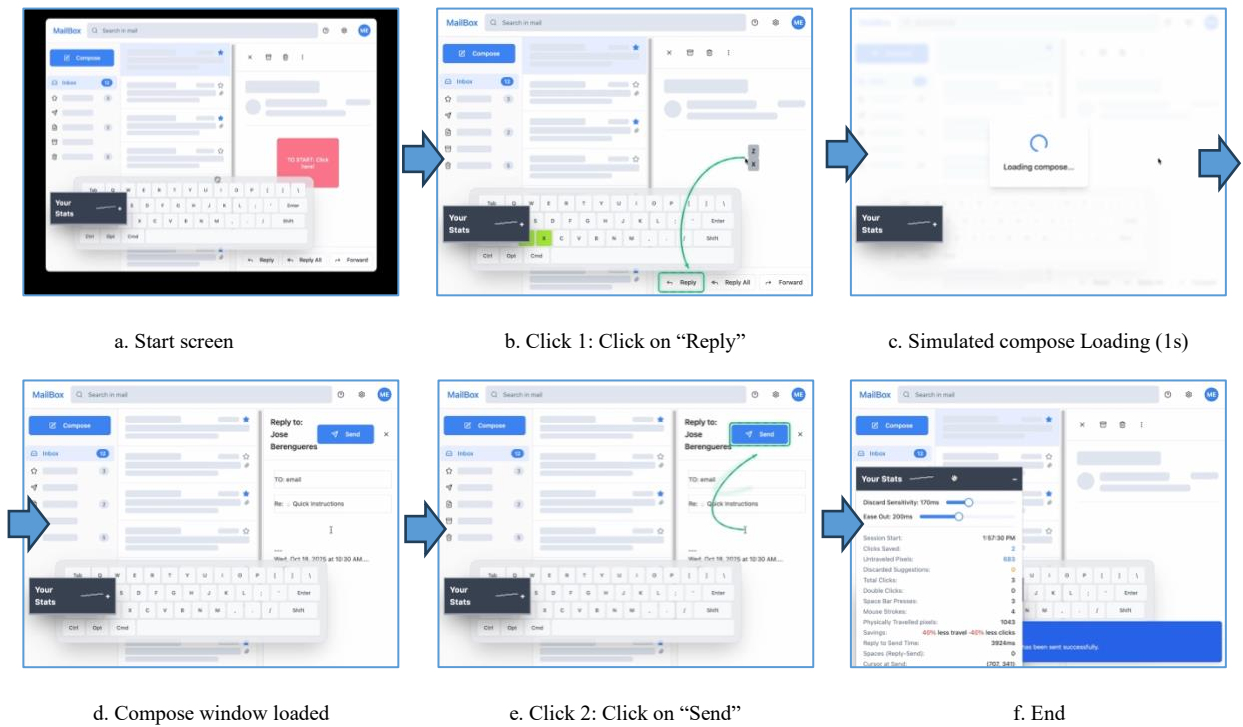

a. Start screen    b. Click 1: Click on "Reply"    c. Simulated compose Loading (1s)

d. Compose window loaded    e. Click 2: Click on "Send"    f. End

Figure 3: An familiar email mockup used to test feasibility. Here screenshots intentionally not at full screen for figure visualization purposes.

### 4.2 Interaction Logic

The prototype introduces a universal chord shortcut composed of three keys:

1. Z + X hold down → enters PAD mode and previews the system's best guess.
2. Z + X hold down + toggle Spacebar → cycles through remaining ranked candidates ($\leq 6$).
3. Z + X release timing → encodes user intent: simultaneous release of Z + X accepts, sequential release discards.

The distinction between sequential and simultaneous release is 170ms by default and user-adjustable in the settings panel. See also Algorithm 1 and Figure 3f.

## 4.3 Visual Feedback

Each predicted target is highlighted through a curved chord connecting the cursor to the target (e.g., *Reply* or *Send* buttons). Users strongly preferred curved connectors over straight ones, consistent with general perceptual findings that curved shapes are judged more pleasant [16]. The reason is unknown. On accept, the chord retracts toward the target; on discard, it retracts away. These transitions use ease-in-out animation [23] (200ms default) to reduce abruptness and maintain continuity between preview states. (See ghost in Figure 2).

## 4.4 Onboarding

To minimize cognitive overload, the prototype includes a progressive onboarding system (not part of PAD). Users first learned only one action—how to preview the next suggestion. After the first N suggestions are exhausted, a contextual cheat-sheet appears inline, revealing the two remaining shortcuts to learn (accept or discard). This guided exposure mirrors shortcut-learning methods used in *KeyMap* [14] and *FingerChord* [20]. However, this information hiding, while clever in terms of preventing overload, also comes with its own drawbacks [21].

## 4.5 Instrumentation and Logging

Through a floating window called *Your Stats*, the prototype logs user actions locally in CSV format, capturing:

1. number of key presses and releases,
2. number of previews and discards,
3. pointer travel distance (in pixels),
4. saved distance per accepted chord.

These metrics support the quantitative comparison with trackpad input.

## 4.6 Technical Deployment

The prototype is a React Native static website that runs entirely in the browser, requiring no installation or external APIs. It is hosted online at https://email-playhouse2.lovable.app and open-sourced at https://github.com/orioli/email-playhouse. The system thus represents a "Wizard-of-Oz" style prototype emulating predictive.

## 4.7 Limitations

As mentioned, all predictions are deterministic and pre-scripted to allow controlled evaluation of the PAD interaction without confounds from model accuracy or latency. Consequently, findings speak to interaction feasibility and user experience, not to algorithmic performance.

## 5 EXPERIEMNTAL METHODOLOGY

The evaluation focuses RSI risk mitigation rather than characterizing the speed performance or accuracy as an input method. Nevertheless, two tests are used, the email mockup to evaluate feasibility and the the ISO-4931-9 test to compare throughput with trackpad.

### 5.1 Email mockup

#### 5.1.1 Objectives

The evaluation aimed to determine whether the proposed **PAD** interaction could

1. (RQ1) reduce physical hand motion compared with a standard mouse or trackpad
2. (RQ2) maintain comparable task-completion time VS. trackpad
3. (RQ3) maintain agency

### 5.1.2 Experimental Setup

A within-subjects laboratory study was conducted using the prototype. The environment simulated an email client with clickable elements such as Reply, Reply All, Send, Archive, and Forward. The first suggestion is always predefined to be correct. The study was approved by the Institutional Ethics Review Board (Reference: 2025-01337-01). All experiments were conducted on a 14-inch MacBook Pro (Apple M4, 16 GB RAM, 1 TB SSD, Nov 2024) running macOS Sequoia 15.6.1. The prototype was executed in full screen in a Google Chrome browser (Version 141.0.7390.123, Official Build, arm64). The display was a built-in Liquid Retina XDR panel with a native resolution of 3024 × 1964 pixels. No external input devices were connected; all interactions were performed using the integrated keyboard and trackpad.

### 5.1.3 Task

Participants completed a **short onboarding phase** (≈ 1–5 minutes) to learn the keyboard shortcuts for PAD. After onboarding, each participant was asked performed a *reply-and-send* task. The task required one click to start the session and two subsequent mouse actions—one click on a *Reply* button in the main window and one click on *Send* in the compose screen. This scenario represented a minimal yet familiar sequence of GUI operations common in everyday use. Figure 3 illustrates the tasks in six steps. Figure 4 shows a user testing the prototype (not part of the study).

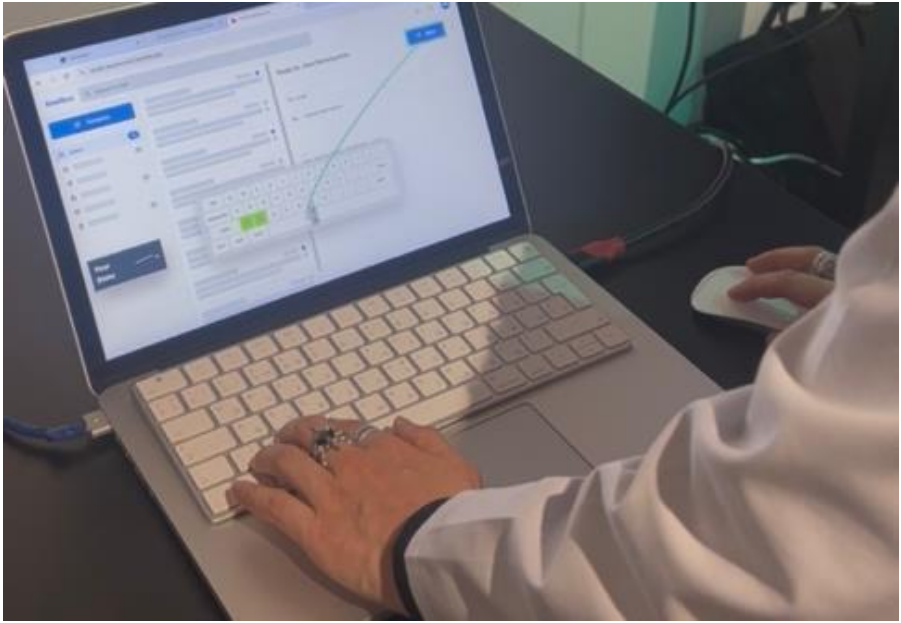

Figure 4: A user from middle-income developing country (not part of the study) tests the online prototype with an aging five-year-old laptop.

### 5.1.4 Participants

Two groups were recruited. To test PAD, 13 volunteers (6 female, 7 male, aged 18–30) participated. All were daily computer users, familiar with shortcut keys but not with chorded input. To act as control group (trackpad) 8 students were recruited (4 female, aged 18–27). Participation was voluntary; no monetary compensation was provided.

### 5.1.5 Measurements

Objective metrics

1. Task-completion time: time from cue onset to target activation
2. Pointer-travel distance: in-browser pixel distance of pointer path (via JavaScript logging)
3. Number of keypresses

4. Number of strokes on trackpad
    5. Number of clicks

Subjective metrics: verbal feedback and written comments.

### 5.2 Throughput under Varying Top-N Accuracy

After evaluating the feasibility we took on the task of comparing to other devices using the ISO-9241-9. To do this we used the same low code app to create an online version of the ISO-9241-9 https://ai-mouse.lovable.app/. We recruited N=26 participants aged 18 to 35, with a balanced gender. And for each, several runs of 22 trials where conducted in different conditions such as varying ID (from 4-6), varying input device trackpad or PAD, varying the top-3 accuracy distribution: and ideal case (@100% with spell checker like distribution of 95% correct first suggestion 4% correct $2^{nd}$ suggestion, 1% correct $3^{rd}$ suggestion,) and a worst case scenario for a 100% accuracy (with uniform distribution $1/3^{rd}$ first suggestion $1/3^{rd}$ second and $1/3^{rd}$ ) where used to compare different accuracies.

## 6 RESULTS

Because of the exploratory nature of the email mockup , no inferential statistics (e.g., confidence intervals or p-values) are reported. The goal of this initial evaluation was to demonstrate interaction **feasibility**, not to quantify performance or accuracy. The analysis therefore focuses on qualitative trends—directional reductions in motion and participants' subjective impressions—sufficient to establish the viability of the PAD to reduce RSI risk and maintain agency.

### 6.1 Email mockup

#### 6.1.1 Motion Reduction

**Figure 5** compares two groups **PAD (ai-predict)** users and the **trackpad** (no AI) users. It shows a box plot comparison across three key metrics. Participants using **PAD** completed both the onboarding phase in (~1-5 minute) and the email "reply-and-send" task (1<minute), with reduced physical motion of the hand (See count of saved clicks). The total number of physical **clicks** effectuated dropped to zero for PAD users during both onboarding and the send email task. The count of finger strokes (trackpad swipes or pointer travel segments) was also substantially lower with PAD even though the trackpad users underwent a shorter onboarding phase as they did not need to learn the shortcut. In terms of speed, we could not determine any significant difference. However, the median travel distance saved by PAD was approximately 3,000 pixels of pointer travel per user for every five accepted chord suggestions on average, corresponding to roughly 600 fewer pixels per replaced click (this includes both onboarding and email task phases).

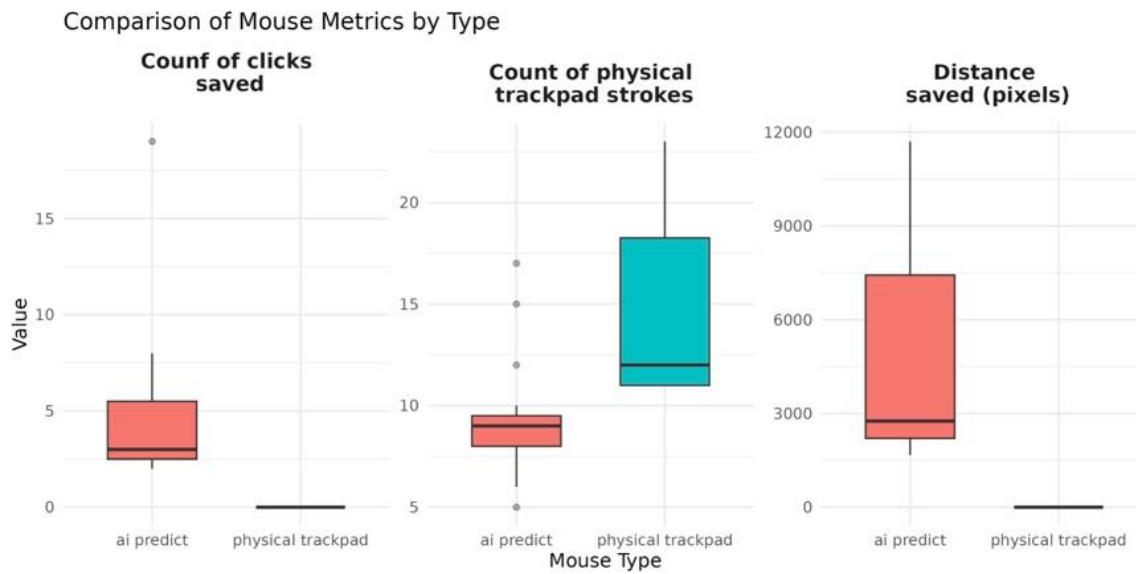

Figure 5: Qualitative comparison of AI-assisted input with chords (ai-predict) and trackpad when carrying a simple email task. *Includes onboarding and the email task.

Participants generally described the interaction as 'fun'. Several "Aha!" moments were logged once the shortcuts were mastered. Several participants (three) commented that once they understood the chord system, it felt playful. Two participants, however, noted that the discard action—releasing only one key to cancel the prediction—took time to master and had difficulty releasing both fingers at the same time (the threshold to determine concurrency of release was set to 170ms). Participants with particularly thick fingers (not part of the study) had more trouble with the release mechanism to Accept-Discard. Three users noted that they found the system useful for **accessibility** to substitute for their aging, non-working trackpads, rather than as a pain relief strategy.

#### 6.1.2 Qualitative Observations

1. **Learning curve:** Mastery required about two minutes on average; users appreciated the minimal grammar. QW keys are less error prone than ZX because they can be pressed more firmly.
2. **Visual feedback:** Curved chord shapes were unanimously preferred over straight ones.
3. **Few errors or miss clicks.**

Overall, the PAD prototype demonstrated it can achieve **substantial** elimination of pointer travel and click without loss of agency or substantial cognitive cost (RQ3). The lack of miss clicks indicates that there is low risk of the user feeling loss of agency.

### 6.2 Evaluation of Top-3 Accuracy

A challenge when simulating predictive system is what to do when the first suggestion is wrong. Here we take cues from spell checkers and use the same metrics: Top-N accuracy metrics. We choose N=3 because most spell checkers work with N=3 and we test for 100% accuracy to understand the best possible performance. Figure 6 shows a normalized time across trial index that seems to stabilize after trial number 5. Therefore, we exclude the first 5 trials from the statistics.

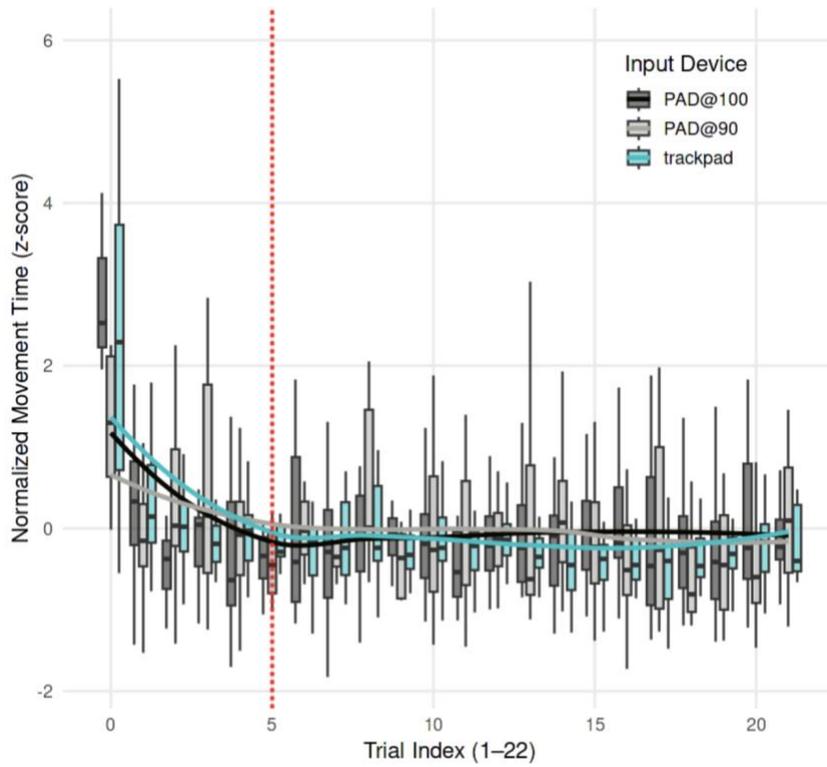

Figure 6: The experimental learning curve was found by normalizing time per trial ordinal. Only trials after 5 are considered.

Figure 7 shows the performance of the three cases. Y-axis is time to complete a trial of the iso task and X is the index of difficulty as log( A/W +1). The lines are linear regressions with 95% CI. In blue the ideal case of PAD @100% with spell checker like distribution of suggestions. The green line corresponds to the trackpad. The red line is PAD@100% with the unform distribution of suggestions (i.e.. The first suggesting is only right 1/3$^{rd}$ of the time, forcing the user to cycle though options). Figure 8 compares the throughput in bps across IDs, lines show 95% CI.

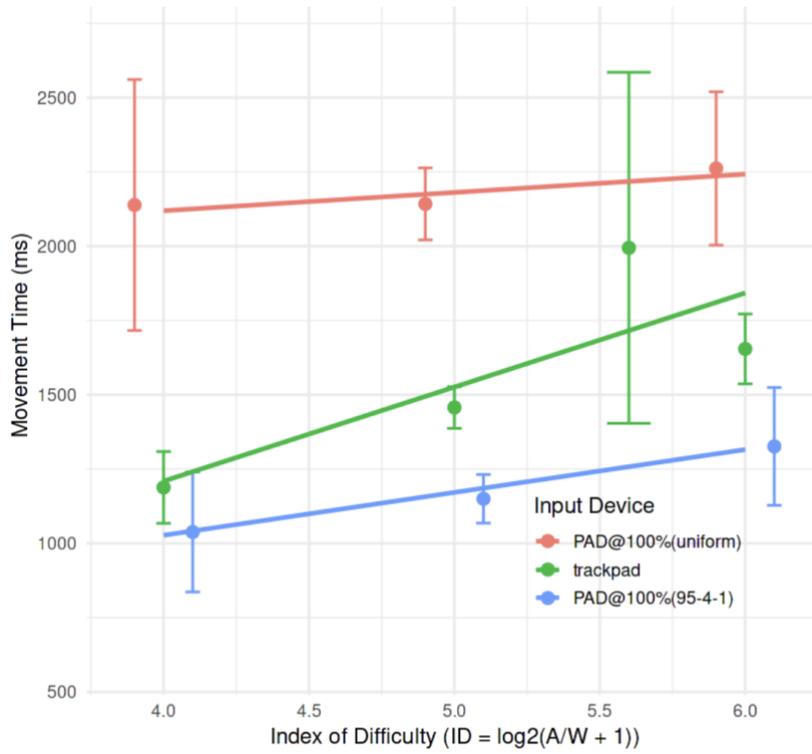

Figure 7: Comparison of trackpad VS. two cases of AI assisted input @Top-3 100% Accuracy. The ideal case for AI-assisted input (bottom) is barely quicker the trackpad (middle). The worse-case for AI-assisted input where the 1$^{st}$ suggestion is only correct the1/3$^{rd}$ of the time is shown at the top. The Trackpad regression line follows Fitts Law, both AI-assisted follows a hybrid Hicks-Fitts models with less slope.

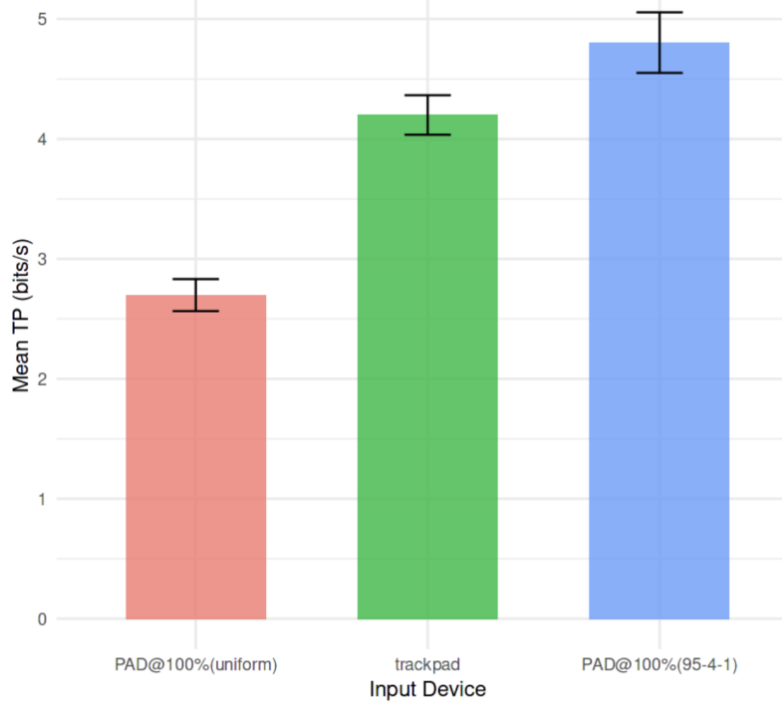

Figure 8: Comparison of throughput. The gain between the trackpad and the ideal case for AI assisted input is marginal (<0.5bits/s).

## 6.3 Motion-Exposure Reduction

Similarly, Figure 9 compares the mean number of strokes per trial. Figure 10 shows the error rate, where the ideal case of PAD@100% shows a significant reduction in errors. Users miss one in 11 clicks with the trackpad and one in 20 with PAD@100% uniform distribution caseand nearly zero in the ideal case. Table 1 summarizes the key metrics of the charts.

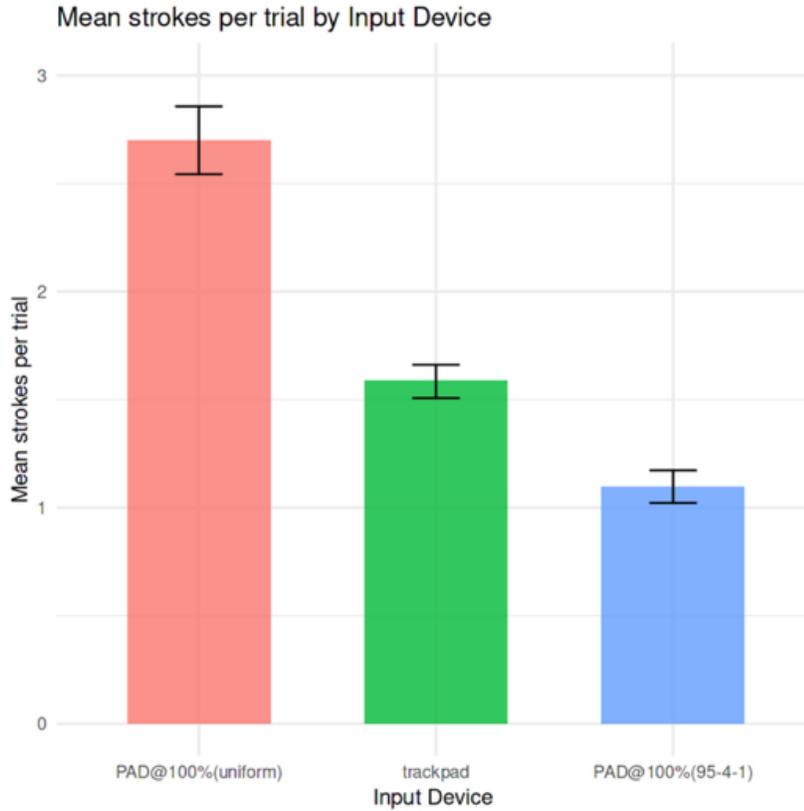

Figure 9: Comparison of Strokes of keyboard or trackpad "segments" per trial.

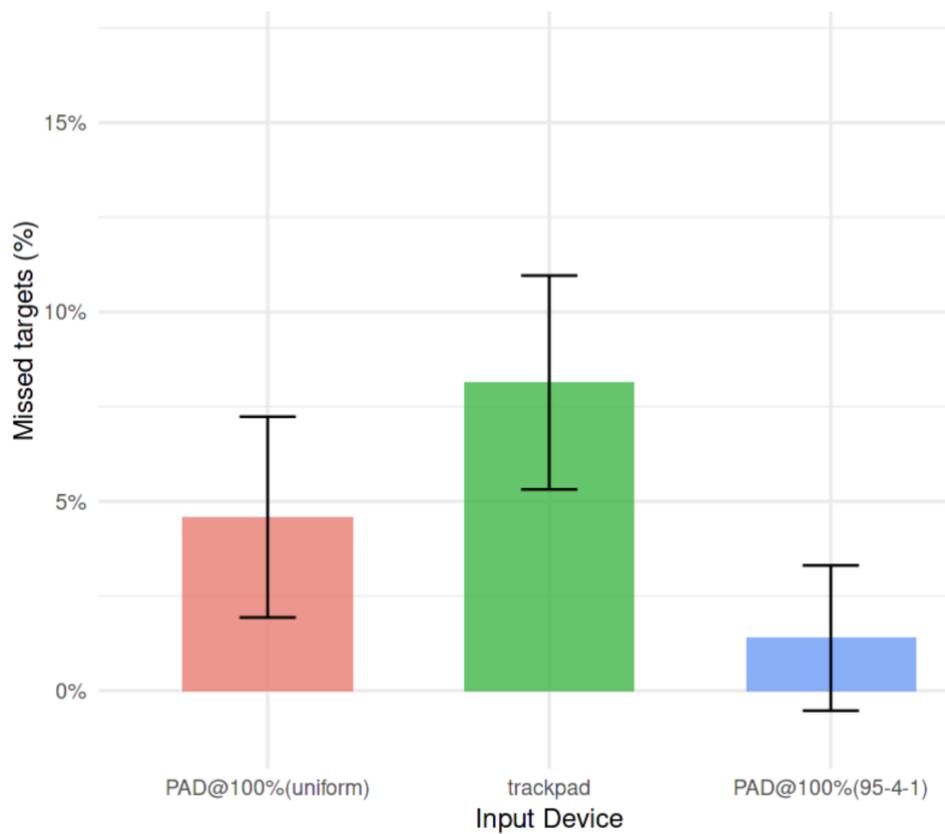

Figure 10: Average error rates as percent of missed targets across IDs 4, 5, 6.

Table 1 Comparing three input methods

| ISO 9241-9 (KP) | PAD@ Top-3 100% Acc (95-4-1) | PAD@ top-3 100% Acc (uniform: 33-33-33) | Trackpad* |
|---|---|---|---|
| **N trials** | 112 | 192 | 310 |
| **Mean stroke (gesture segments or keypresses per trial)** | 1.08 | 2.8 | 1.67 |
| **Min stroke count** | 1 | 1 | 1 |
| **Max stroke count** | 5 | 9 | 7 |
| **mean TP 95% CI (bps)** | 4.8 [4.5,5.1] | 2.7 [2.6,2.8] | 4.2 [4.0,4.4] |
| **Median TP (bps)** | 4.6 | 2.5 | 3.6 |
| **SD TP (bps)** | 1.5 | 1.1 | 2 |

# 7 DISCUSSION

## 7.1 Motion Savings

The Preview–Accept–Discard (PAD) paradigm reframes graphical interaction from a movement problem to a decision problem. By shifting part of the pointing workload from fine-motor control (Fitts's Law) to cognitive selection (Hick's Law), PAD aims to reduce physical motion without increasing task time. However, as predicted by hybrid Hicks–Fitts models, complete separation of the two processes is not possible. Even with ideal predictive accuracy, users still engage in micro-adjustments and visual verification that preserve Fitts-like feedback loops. Consequently, PAD rarely exceeds the speed of a trackpad except under near-perfect predictive conditions, yet it achieves **large reductions** in total hand motion (about 600px of untraveled pixels on the screen per click in our email mockup setup).

## 7.2 Agency in Predictive Interfaces

Automation in mixed-initiative systems can easily erode user agency through over-trust or under-scrutiny. PAD mitigates this risk by embedding explicit acceptance timing within the interaction loop: the user remains in control of whether to accept, cycle, or discard a prediction at every step. Participant feedback indicated that the visibility of the on-screen chords enhanced situational awareness and reinforced a feeling of control.

These findings align with Amershi et al.'s guidelines for human–AI interaction and Eiband et al.'s emphasis on transparency and feed-forward cues as prerequisites for trust [13].

## 7.3 Adoption and Learnability

Shortcut-based systems often fail to gain adoption because users forget mappings or never perceive the benefits [26, 38]. PAD addresses this by coupling prediction with contextual feedback: users preview likely actions before committing, eliminating the need to memorize key combinations. Participants mastered the core grammar—preview, cycle, accept/discard—within minutes in the email mockup, confirming its low learning threshold. Those with larger fingertips reported some difficulty in performing simultaneous key releases, suggesting that future implementations could adjust the release-timing window adaptively or provide calibration. In line with the novice-to-expert transition models proposed by Bailly et al. [4] and Lewis and Vogel [24], PAD's guided discoverability supports progressive skill acquisition without imposing recall burdens.

## 7.4 Inclusive and Sustainable Computing

Beyond ergonomics, PAD might support accessibility and sustainability goals. By reducing dependence on the trackpads (which cost $100s to repair) and mice ($20s), PAD can make everyday computing more inclusive for users with limited dexterity, tremor, or chronic pain—groups often underserved by traditional pointing devices. In many educational and developing economic contexts, the trackpads of older laptops fail long before the CPU or display, often due to mechanical wear or delamination. PAD effectively extends the usable lifespan of aging laptops whose pointing hardware has degraded. See also Figure 4 (right) where the user is carrying an extra mouse as the trackpad of her aging laptop has stopped working. Note also the Bluetooth keyboard covering the built-in keyboard for similar reasons. This not only improves accessibility but also reduces electronic waste.

## 7.5 Limitations and Future Work

The current prototype used **hard-coded predictions**, a constrained task domain (email client) and a standard ISO task initially developed for non-keyboard tasks. It relies on an assumption that the keyboard use is less likely to cause RSI. This assumption has not been proven in a clinical trial yet. Future studies can:

1. Build a AI training dataset of point and click tasks with HTML DOM level information
2. Quantify fatigue reduction via EMG or wrist-motion sensors
3. Investigate longitudinal medical outcomes

## 8 CONCLUSION

A standard mouse switch such as the Omron D2FC-F-7N is typically rated for about 20 million actuations, while a mechanical Cherry-type keyboard switch may endure 50–100 million before failure. More complex mechanisms—like professional-grade DSLR shutters—are rated for 300,000–1,000,000 cycles, and even a DVD tray-eject mechanism seldom exceeds 50,000 operations. By comparison, an average computer user performs over 30 million clicks during a career—a workload rivaling or exceeding the endurance of many mechanical systems. This imbalance underscores that the prevalence of repetitive strain injury (RSI) reflects biomechanical limits of the human hand, not merely poor device design.

By removing motion rather than merely optimizing it, AI-assisted predictive input methods such as PAD offer a new ergonomic pathway. They can potentially extend the usable life of aging laptops by reducing reliance on fragile trackpads and promote inclusion for users with limited dexterity or chronic strain. However, since carpal tunnel incidence in the left and right hands differs only marginally, further clinical studies are needed to verify the assumed ergonomic benefits. Moreover, PAD and similar systems achieve speed gains only under ideal predictive accuracy: as hybrid Hicks–Fitts models suggest, the cognitive cost of verifying a prediction can equal the motor cost of pointing. Ultimately, the most productive application of AI may not be at the menu-selection level but at the application level, where reducing the overall number of required interactions could yield more meaningful and sustainable ergonomic gains.

**ACKNOWLEDGMENTS**

We gratefully acknowledge Dina Kulina (NUSOM), Joe Hutsko for insights on carpal tunnel syndrome. Jörg Conrad (KTH); James Benedict (UMA), Håkan Hansson (UMA), and Kent Brodin (UMA); Gustav Eckerbom (KTH), Waldemar Jenek (Monash), and Dhanya Menoth Mohan (Monash) helped with 3D-printing prototypes; Linda Kan (KTH) offered long-term user feedback; and Karl Meinke (KTH) provided financial support for ethics board review. We also acknowledge Jack Lo (Evoluent), Simon Lancaster (Omni Ventures), Heiko Drewes (LMU), Donn Koh (NUS), and John Van Hooft (BakkerElkhuizen) for key insights on ergonomics and Timur Umurzakov (NU).